\documentclass{webofc}
\usepackage[varg]{txfonts}

\usepackage{comment}

\usepackage[colorlinks=true, citecolor=blue, linkcolor=blue, urlcolor=blue, pdfborder={0 0 0}, breaklinks=true]{hyperref} 
\renewcommand{\url}[1]{\href{#1}{#1}}
\renewcommand{\doi}[1]{\url{https://doi.org/#1}}
\newcommand{\arxiv}[1]{\url{https://arxiv.org/abs/#1}}

\DeclareFontFamily{OT1}{pzc}{}
\DeclareFontShape{OT1}{pzc}{m}{it}{<-> s * [1.10] pzcmi7t}{}
\DeclareMathAlphabet{\mathpzc}{OT1}{pzc}{m}{it} 

\usepackage{tikz}
\usetikzlibrary{cd}
\usepackage{relsize} 

\usepackage{lineno}
\newcommand*\patchAmsMathEnvironmentForLineno[1]{
  \expandafter\let\csname old#1\expandafter\endcsname\csname #1\endcsname
  \expandafter\let\csname oldend#1\expandafter\endcsname\csname end#1\endcsname
  \renewenvironment{#1}
     {\linenomath\csname old#1\endcsname}
     {\csname oldend#1\endcsname\endlinenomath}}
\newcommand*\patchBothAmsMathEnvironmentsForLineno[1]{
  \patchAmsMathEnvironmentForLineno{#1}
  \patchAmsMathEnvironmentForLineno{#1*}}
\AtBeginDocument{
\patchBothAmsMathEnvironmentsForLineno{equation}
\patchBothAmsMathEnvironmentsForLineno{align}
\patchBothAmsMathEnvironmentsForLineno{flalign}
\patchBothAmsMathEnvironmentsForLineno{alignat}
\patchBothAmsMathEnvironmentsForLineno{gather}
\patchBothAmsMathEnvironmentsForLineno{multline}
}

\begin{document}

\title{\vspace*{-0.2cm} 
Optimising HEP parameter fits via Monte Carlo\\
weight derivative regression} 

\author{
\firstname{Andrea}
\lastname{Valassi}\inst{1}
\fnsep\thanks{
  \email{andrea.valassi@cern.ch}
}}

\institute{
CERN, Information Technology Department, 
CH-1211 Geneva 23, Switzerland 
}

\abstract{
HEP event selection is traditionally considered 
a binary classification problem, involving 
the dichotomous categories of signal and background. 
In distribution fits for particle masses or couplings, 
however, signal events are not all equivalent, 
as the signal differential cross section 
has different sensitivities to the measured parameter 
in different regions of phase space. 
In this paper, 
I describe a mathematical framework for 
the evaluation and optimization of HEP parameter fits, 
where this sensitivity is defined 
on an event-by-event basis, 
and for MC events it is modeled in terms 
of their MC weight derivatives 
with respect to the measured parameter. 
Minimising the statistical error on a measurement 
implies the need to resolve (i.e. separate) 
events with different sensitivities, 
which ultimately represents 
a non-dichotomous classification problem. 
Since MC weight derivatives 
are not available for real data, 
the practical strategy I suggest consists 
in training a regressor of weight derivatives 
against MC events, and then using it 
as an optimal partitioning variable 
for 1-dimensional fits of data events. 
This CHEP2019 paper is an extension of the study 
presented at CHEP2018: 
in particular, 
event-by-event sensitivities 
allow the exact 
computation of the ``FIP'' ratio 
between the Fisher information
obtained from an analysis 
and the maximum information 
that could possibly be obtained 
with an ideal detector. 
Using this expression, I discuss 
the relationship between FIP and two metrics 
commonly used in Meteorology (Brier score and MSE), 
and the importance of ``sharpness'' 
both in HEP and in that domain.
I finally point out that HEP distribution fits 
should be optimized and evaluated 
using probabilistic metrics (like FIP or MSE), 
whereas ranking metrics (like AUC) 
or threshold metrics (like accuracy) 
are of limited relevance for these specific problems.
}

\maketitle

\newcommand{\ytru}{\theta}
\newcommand{\ytruref}{{\ytru_\text{ref}}}
\newcommand{\Mref}{{M_\text{ref}}}
\newcommand{\Var}{\mathrm{var}}
\newcommand{\yhat}{\hat{\ytru}}
\newcommand{\sigs}{\sigma\!_s}
\newcommand{\sigsk}{\sigma\!_{s,\kbin}}
\newcommand{\sigsref}{\sigma\!_{s,\text{ref}}}
\newcommand{\ofmeas}{^{\mathrm{(meas)}}}
\newcommand{\nmeas}{M}
\newcommand{\lumi}{\mathcal{L}}
\newcommand{\If}{\mathcal{I}}
\newcommand{\Ify}{\If_\ytru}
\newcommand{\Ifs}{\If_{\sigs}}
\newcommand{\Dy}{\Delta\ytru}
\newcommand{\Ds}{\Delta\sigs}

\newcommand{\ievt}{i}

\newcommand{\wi}{w_\ievt}

\newcommand{\gvar}{\gamma}
\newcommand{\gi}{\gvar_\ievt}
\newcommand{\gbar}{\langle\gvar\rangle}
\newcommand{\gbark}{\gbar_\kbin}

\newcommand{\ddvar}{\delta}
\newcommand{\ddi}{\ddvar_\ievt}
\newcommand{\ddbar}{\langle\ddvar\rangle}

\newcommand{\ideal}{^{\smash{\text{(ideal})}}}
\newcommand{\idealALL}{^{\smash{\text{(ideal,\,$\sALL$})}}}
\newcommand{\true}{^{\smash{\text{(true})}}}
\newcommand{\mmin}{^{\smash{\text{(min})}}}
\newcommand{\mmax}{^{\smash{\text{(max})}}}

\newcommand{\xvar}{x}
\newcommand{\xvec}{\mathbf{\xvar}}
\newcommand{\xveci}{\xvec_\ievt}
\newcommand{\xvecgen}{\xvec\true}
\newcommand{\xvecgeni}{\xvecgen_\ievt}

\newcommand{\fip}{\text{FIP}}
\newcommand{\fipo}{\text{FIP}_1}
\newcommand{\fipt}{\text{FIP}_2}
\newcommand{\fiph}{\text{FIP}_3}
\newcommand{\fipsel}{\text{FIP}_{\text{sel}}}
\newcommand{\fipbin}{\text{FIP}_{\text{part}}}

\newcommand{\greg}{\mathpzc{q}} 
\newcommand{\gregi}{\greg_\ievt}
\newcommand{\gregxi}{\greg(\xvec_\ievt)}
\newcommand{\gregx}{\greg(\xvec)}
\newcommand{\gregk}{\greg_{(\kbin)}}

\newcommand{\MSE}{{\text{MSE}}}
\newcommand{\MSEk}{{\text{MSE}_{(\kbin)}}}

\newcommand{\dvar}{\mathcal{D}}
\newcommand{\dthr}{\dvar_{\mathrm{thr}}}

\newcommand{\phis}{\phi}
\newcommand{\phisk}{\phis_{\kbin}}

\newcommand{\nbin}{K}
\newcommand{\kbin}{k}
\newcommand{\kobin}{{k_1}}
\newcommand{\ktbin}{{k_2}}
\newcommand{\Kbin}{\mathcal{K}}
\newcommand{\sumks}{\sum_\kbin\!}
\newcommand{\sumkk}{\sum_{\kbin\!=\!1}^K}

\newcommand{\ntot}{N_{\mathrm{tot}}}
\newcommand{\nsel}{N_{\mathrm{sel}}}
\newcommand{\nvar}{N}
\newcommand{\nrej}{N_{\mathrm{rej}}}
\newcommand{\stot}{S_{\!\mathrm{tot}}}
\newcommand{\ssel}{S_{\!\mathrm{sel}}}
\newcommand{\srej}{S_{\!\mathrm{rej}}}
\newcommand{\btot}{B_{\mathrm{tot}}}
\newcommand{\bsel}{B_{\mathrm{sel}}}
\newcommand{\brej}{B_{\mathrm{rej}}}
\newcommand{\sALL}{S_{\!\mathrm{ALL}}}

\newcommand{\eff}{\epsilon}
\newcommand{\effs}{\epsilon_s}
\newcommand{\effb}{\epsilon_b}
\newcommand{\gpur}{\varrho} 
\newcommand{\pur}{\rho} 
\newcommand{\purk}{\pur_\kbin}
\newcommand{\puro}{\pur_1}
\newcommand{\purt}{\pur_2}
\newcommand{\effks}{\eff_{s,\kbin}}
\newcommand{\effkb}{\eff_{b,\kbin}}

\newcommand{\ssetk}{s_\kbin}  
\newcommand{\bsetk}{b_\kbin}  
\newcommand{\nsetk}{n_\kbin}  
\newcommand{\nset}[1]{n_{#1}} 
\newcommand{\sset}[1]{s_{#1}} 

\newcommand{\nsetx}{n(\xvec)}
\newcommand{\ssetx}{s(\xvec)}
\newcommand{\bsetx}{b(\xvec)}
\newcommand{\purx}{\pur(\xvec)}
\newcommand{\phisx}{\phis(\xvec)}
\newcommand{\gbarx}{\gbar(\xvec)}
\newcommand{\ddbarx}{\ddbar(\xvec)}
\newcommand{\ltru}{\lambda}
\newcommand{\ltx}{\ltru(\xvec)}
\newcommand{\dx}{d\xvec}

\newcommand{\gbarxi}{\gbar(\xveci)}

\newcommand{\ddbark}{\ddbar_\kbin}

\newcommand{\mselk}{m_\kbin}
\newcommand{\mselko}{m_\kobin}
\newcommand{\mselkt}{m_\ktbin}
\newcommand{\mselv}{{\mathbf m}}
\newcommand{\mself}{m_1}
\newcommand{\msell}{m_\nbin}

\newcommand{\fpkpy}[1]{\frac{\partial#1}{\partial\ytru}}
\newcommand{\pkpy}[1]{\partial#1/\partial\ytru}
\newcommand{\fsenky}[1]{\frac{1}{#1}\fpkpy{#1}}
\newcommand{\senky}[1]{(1/#1)(\pkpy{#1})}
\newcommand{\senkys}[1]{(1/#1)^2(\pkpy{#1})^2}

\newcommand{\fpxkpy}{\fpkpy{\lsigsk}}
\newcommand{\pxkpy}{\pkpy{\lsigsk}}
\newcommand{\fsenxky}{\fsenky{\lsigsk}}
\newcommand{\senxky}{\senky{\lsigsk}}

\newcommand{\Prob}[1]{\text{Prob}_{(#1)}}

\newcommand{\sumik}{\sum_{\ievt\in\kbin}}
\newcommand{\sumiksig}{\sum_{\ievt\in\kbin}^{\text{(Sig)}}}

\section{Introduction}
\label{sec:intro}

The point estimation of physics parameters,
such as the measurement of a cross~\mbox{section}
or of a particle's mass or couplings,
is an important category of data analysis problems
in experimental High Energy Physics (HEP).
Optimizing these measurements
ultimately consists in minimizing the 
combined statistical and systematic 
errors on the measured \mbox{parameters}.
In this paper, 
I only discuss
the minimization of 
the statistical error $\Dy$,
in the measurement of a single parameter~$\ytru$
from the binned fit 
of a multi-dimensional distribution 
of selected events.
This 
implies the optimization
of two analysis handles:
event selection,
i.e. the~\mbox{criteria} 
for signal-background discrimination,
and event partitioning,
i.e. the choice of binning variables. 

This article follows up on that I presented
at CHEP2018~\cite{bib:chep2018av}.
As in that occasion,
two central points of my study
are a discussion of 
evaluation and training metrics for 
the data analysis tools used in the measurement,
and a comparison of these metrics to those 
used in other scientific domains.
The starting point of this analysis is, again,
the calculation of 
the statistical error $\Dy$
in a binned fit for the parameter~$\ytru$
and its comparison to
the minimum error $\Dy\ideal$ which could be achieved
in an ``ideal'' case.
Minimizing $\Dy$ is 
equivalent to maximizing 
$(\Dy\ideal)^2/(\Dy)^2$, a metric in [0,1]
that I refer to as ``Fisher Information Part''~(FIP).

This research differs from and extends my 
CHEP2018 work in two respects.
First, 
it shifts the focus
from event selection, 
which is a binary classification problem,
to event partitioning,
and it shows that the latter can be 
addressed as a non-binary regression problem.
The key improvement 
is the derivation of~$\Dy$
in terms of the event-by-event sensitivity $\gi$ 
of each event $\ievt$ to the parameter $\ytru$,
rather than in terms of the
\mbox{bin-by-bin sensitivity in bin~$\kbin$}
(which is simply the average
event-by-event sensitivity $\gbar_{\kbin}$ in that bin).
I show that 
the optimal partitioning strategy
consists in binning events
according to their sensitivity $\gi$,
and~I use this to derive
the minimum error $\Dy\ideal$ achievable 
with an ideal detector 
and an ideal analysis method.
While $\gi$ can be computed for Monte Carlo (MC) events
from the derivative of their MC weight with respect to $\ytru$,
however, $\gi$ 
is not available for real data events:
the practical strategy I suggest consists 
in training a regressor~$\gregi$ 
of $\gi$ on MC events,
and using it as an optimal partitioning variable 
for a 1-dimensional fit of data events. 
The FIP metric can be used
both for evaluating the quality of the result,
and as a loss function for training the regressor~$\gregi$.
In this context,
where only statistical errors are considered,
event partitioning can be seen 
as a generalization of event selection,
which is a simpler, binary, sub-case.
Rather than simply separating 
signal events, which are sensitive to $\ytru$,
from background events, which are not,
the problem to address is how to 
resolve, i.e. separate,
events with different sensitivities to $\ytru$: 
this ultimately represents 
a non-dichotomous classification problem.

The second new contribution
of this research is the comparison
to other non-HEP scientific domains,
beyond those I had previously considered. 
In my CHEP2018 study,
I had mainly considered 
the evaluation metrics
for binary classification problems
in Medical Diagnostics, 
Information Retrieval,
and Machine Learning research.
I had also briefly discussed
a few metrics used in those fields
to go beyond a strictly dichotomous categorization 
of the true event categories,
or to take into account the ranking of events 
when a scoring classifier
is used instead of a binary discrete classifier.
In this paper, 
I extend this comparative analysis by
pointing out 
the close relationship between FIP and two metrics 
commonly used in Meteorology 
(the ``Brier score''~\cite{bib:brier1950}
and the ``Mean Squared Error'' or MSE), 
and the importance of ``sharpness'' 
both in HEP and in that domain.
More generally, I suggest
that HEP distribution fits 
should be optimized and evaluated 
using probabilistic metrics
(like MSE, or FIP)
as is commonly the case in Meteorology and Medical Prognostics,
whereas ranking metrics 
(like the ``Area Under the ROC Curve'' or AUC) 
or threshold metrics (like ``accuracy''),
which are widely used in Medical Diagnostics,
are of limited relevance for these specific problems.

The outline of this paper is the following.
Section~\ref{sec:fits} 
describes a mathematical \mbox{framework}
for discussing statistical error minimizations
in HEP parameter fits, and 
the use of MC weight derivative regression
to optimize event partitioning.
It also discusses
the relationship
between FIP and MSE as training metrics
for Decision Tree regressors,
using a decomposition of MSE
into calibration and sharpness
that is copied from Meteorology.
Section~\ref{sec:brier}
points out 
the relevance of probabilistic metrics,
more than threshold or ranking metrics, 
in both HEP and Meteorology.
An outlook for this research
and some conclusions are given 
in Section~\ref{sec:concl}.

\section{Statistical errors in HEP binned fits of a parameter \texorpdfstring{$\pmb\ytru$}{theta}}
\label{sec:fits}

\newcommand{\nzelk}{n_{\kbin}}
\newcommand{\nzelko}{n_{\kobin}}
\newcommand{\nzelkt}{n_{\ktbin}}

\newcommand{\ofy}{(\ytru)}
\newcommand{\ofyref}{(\ytruref)}
\newcommand{\ofyder}{(\ytruder)}
\newcommand{\weighti}{\wi}
\newcommand{\sensiy}{\partial\weighti/\partial\ytru}
\newcommand{\ME}{\mathcal{M}}
\newcommand{\MEsq}{|\ME|^2}
\newcommand{\MEsqpar}[1]{|\ME(#1)|^2}

\newcommand{\ytruder}{{\ytru_\If}}

Binned fits for a HEP parameter~$\ytru$ 
rely on splitting all selected events
into $\nbin$ disjoint partitions, or ``bins'',
according to the values of one or more variables
that are
computed as~functions
of the observed properties $\xveci$ of each event~$\ievt$.
When only statistical errors are considered,
the Fisher information $\Ify$ 
about~$\ytru$
which is gained from its measurement,
i.e. the inverse~of~the square 
of the statistical error~$\Dy$,
is easily shown~\cite{bib:chep2018av}
to be the sum of the information contributions
from the independent, and a fortiori uncorrelated,
measurements of~$\ytru$
in these $\nbin$~bins,
\begin{equation}
\setlength{\abovedisplayskip}{\abovedisplayskip-1.2mm}
\setlength{\belowdisplayskip}{\belowdisplayskip-0.7mm}
\Ify
= \frac{1}{(\Dy)^2}
= \sumkk \frac{1}{(\Dy)_\kbin^2}
= \sumkk \nzelk \left( \frac{1}{\nzelk} \fpkpy{\nzelk} \right)^{2} 
\, .
\label{eq:infoyksum}
\end{equation}
where $\nsetk\ofy\!=\!\ssetk\ofy\!+\!\bsetk$ is 
the number of selected events in bin $\kbin$.
This is the sum of 
the number of signal events $\ssetk$, 
which depends on $\ytru$,
and that of background~events $\bsetk$, 
which does not.

\paragraph{MC reweighting and event-by-event sensitivities}
In practice, 
HEP fits of a parameter $\ytru$ 
rely on the theoretical prediction of 
the number of signal events $\ssetk\ofy$ 
in bin~$\kbin$ 
as a function of~$\ytru$,
obtained through MC simulations.
A relatively standard practice 
to derive $\ssetk\ofy$ 
is the MC reweighting technique,
which, for instance,
was used extensively by the LEP experiments
in the late~1990s, 
for measurements 
of both particle masses~\cite{bib:opal-ww-1998,bib:aleph172}
and particle couplings~\cite{bib:aleph-lemaitre-2000}.
This technique is also applicable
to hadron colliders~\cite{bib:gainer2014},
where it has been shown that is
generally feasible also at NLO accuracy~\cite{bib:mattelaer2016}:
it has been pointed out~\cite{bib:gainer2014},
in particular, that it is conceptually and practically simpler
than the Matrix Element Method~\cite{bib:kondo1988,bib:dalitz1992},
which has been extensively used 
at hadron colliders~\cite{bib:me-dz2004,bib:kondo2006,bib:me-cdf2007},
because it does not imply 
the time-consuming integration
over undetermined momenta
which is necessary in that method,
and which can be performed by tools such as MadWeight~\cite{bib:madweight2008}.

Monte Carlo reweighting essentially
consists in the following three steps.
First, a sample 
of MC events for the signal process
is generated
at a reference value $\ytruref$ 
of the parameter~$\ytru$,
and a weight $\weighti\ofyref$ is assigned
to each event $\ievt$;
if unweighted events are generated, they all
have the same $\weighti\ofyref$,
but this is not strictly needed.
Second,
generator-level events 
are passed
through full detector simulation.
Third, 
each detector-level event $\ievt$ is assigned
a weight $\weighti\ofy$ 
at another value of the parameter~$\ytru$;
this is done
by rescaling $\weighti\ofyref$
by the ratio 
of the predicted probabilities 
for $\ytru$ and $\ytruref$
of event $\ievt$,
as described by its 
MC truth (generator-level) properties $\xvecgeni$.
The probability ratio
is typically just a ratio 
of squared matrix elements,
\begin{equation}
\setlength{\abovedisplayskip}{\abovedisplayskip-1.0mm}
\setlength{\belowdisplayskip}{\belowdisplayskip-0.7mm}
\weighti\ofy 
= \left(\frac{\Prob{\ytru}(\xvecgeni)}{\,\Prob{\ytruref}(\xvecgeni)\,}\right)
\,\weighti\ofyref
= \left(\frac{\MEsqpar{\ytru,\xvecgeni}}{\,\MEsqpar{\ytruref,\xvecgeni}\,}\right)
\,\weighti\ofyref
\, .
\label{eq:weighti2}
\end{equation}

The above description applies to signal MC events, but
each background MC event is also assigned a weight $\wi$,
with the important difference that,
by definition, it does not depend on~$\ytru$.
Assuming that all weights $\wi$
take into account 
a normalization factor
to the luminosity of the data, 
the expected number 
of selected signal and background events 
$\nsetk\ofy$ in bin $\kbin$,
as a function of $\ytru$,
can be written
as the sum of the event weights $\weighti$ 
for all MC events $\ievt$ in bin~$\kbin$,
\begin{equation}
\setlength{\abovedisplayskip}{\abovedisplayskip-1.5mm}
\setlength{\belowdisplayskip}{\belowdisplayskip-1.0mm}
\nsetk(\ytru) 
= \sum_{\ievt\in\kbin} \wi(\ytru)
= \sum_{\ievt\in\kbin}^{\text{Sig}} \wi(\ytru)
+ \sum_{\ievt\in\kbin}^{\text{Bkg}} \wi
= \ssetk(\ytru) + \bsetk \, .
\label{eq:nsetksum}
\end{equation}
The bin-by-bin sensitivity of $\nsetk$ to $\ytru$
which appears in Eq.~\ref{eq:infoyksum}
can then be written as
\begin{equation}
\setlength{\abovedisplayskip}{\abovedisplayskip-0.8mm}
\setlength{\belowdisplayskip}{\belowdisplayskip-0.5mm}
\displaystyle
\fsenky{\nsetk}
= \frac{\sumik \fpkpy{\wi}}{\sumik \wi} 
= \frac{\sumik \wi (\fsenky{\wi})}{\sumik \wi}
= \frac{\sumik \wi \gi}{\sumik \wi} 
= \gbar_\kbin \, ,
\,
\label{eq:binbybin}
\end{equation}
i.e. as the weighted average
over all MC events $\ievt$ in bin~$\kbin$,
of the event-by-event sensitivity
\begin{equation}
\setlength{\abovedisplayskip}{\abovedisplayskip-1.1mm}
\setlength{\belowdisplayskip}{\belowdisplayskip-1.0mm}
\gi=\left.\fsenky{\wi}\right|_\ytruder \, .
\label{eq:gi}
\end{equation}
Note that all $\gi$ (and hence $\Ify$)
depend on the value $\ytruder$ of $\ytru$
where $\wi$ and $\pkpy{\wi}$ are~computed
(typically, $\ytruref$).
In a given binning scheme,
the information $\Ify$ of Eq.~\ref{eq:infoyksum}
can then be written~as
\begin{equation}
\setlength{\abovedisplayskip}{\abovedisplayskip-1.7mm}
\setlength{\belowdisplayskip}{\belowdisplayskip-4.0mm}
\Ify
= \sumkk \nsetk \left( \fsenky{\nsetk} \right)^{2}
= \sumkk \nsetk \gbar_\kbin^2 \, .
\label{eq:infoyksumg}
\end{equation}

\paragraph{Beyond the signal-background dichotomy}

For individual signal events $\ievt$,
the event-by-event sensitivity $\gi$
may be positive or \mbox{negative},
and the absolute value of $\gi$ 
may also be significantly different
from one event to another.
Background events, conversely,
all have a zero event-by-event-sensitivity,
because these events, by definition,
are produced by processes 
that are insensitive to the parameter $\ytru$:
\begin{equation}
\setlength{\abovedisplayskip}{\abovedisplayskip-2mm}
\setlength{\belowdisplayskip}{\belowdisplayskip-2mm}
\hspace*{-0.4cm}
\begin{array}{lr}
\displaystyle
\gi \! = \!\left(\fsenky{\wi}\right) 
\!\in\! [-\infty,\!+\infty] \, , \,
\hspace*{0.1cm} \text{if}\,\ievt\in\{\text{Signal}\}
\, ;
\hspace*{0.5cm}
&
\displaystyle
\gi \! = \!\left(\fsenky{\wi}\right) 
\!=\! 0 \, , \,
\hspace*{0.1cm} \text{if}\,\ievt\in\{\text{Background}\} \, .
\end{array}
\hspace*{-0.3cm}
\label{eq:gisb}
\end{equation}
Equation~\ref{eq:infoyksumg} shows 
that the largest contributions 
to the information $\Ify$
come from the~bins~with the largest
average event-by-event sensitivities.
As discussed more in detail later on,
a good measurement is therefore one 
satisfying two criteria:
first, the 
event selection accepts 
the events with sensitivities $\gi$
that are significantly different from zero,
whether positive or negative,
i.e. those with high absolute values of $\gi$
(in the following I will 
refer to these~as events with high sensitivities,
but it should be implicitly understood 
that I refer 
to their~absolute values);
second, the 
event partitioning
resolves events with very different sensitivities
into separate bins,
as it is the average bin-by-bin sensitivity
that determines the contribution to~$\Ify$.

As an example, consider
the measurement of a particle mass $M$
from the fit to the distribution of the 
invariant mass $m$ 
of the decay products of that particle.
The 
sensitivity $\gi$~to~$M$~is positive for the signal events 
on the right of the mass peak ($m\!>\!M$)
and negative for those on its left ($m\!<\!M$).
The events with the highest 
sensitivity (in absolute value) 
are those on the steep 
ascending and descending slopes
to the left and to the right 
of the peak.
Conversely,
the events below the peak or 
on the tails far away from it
have a sensitivity 
that is close to~0.
These low-sensitivity signal events
are not very different~from~background~events,
as the information about $\ytru$
that they provide is 
extremely limited, 
and it is important to separate 
both of these types of events
from high-sensitivity signal events,
so as not to dilute their sensitivity.

In spite of its limitations, 
a 
dichotomous categorization
of events as signal or background is still useful
(especially when considering systematic errors).
Using the symbols
$\purk\!=\!\ssetk/\nsetk$
to indicate the 
selection purity 
and 
$\phisk$ 
to indicate the 
sensitivity of signal events alone
in bin~$\kbin$,
\begin{equation}
\setlength{\abovedisplayskip}{\abovedisplayskip-2mm}
\setlength{\belowdisplayskip}{\belowdisplayskip-2mm}
\phisk 
= \gbar_{\kbin,\text{Sig}} 
= \frac{\sumiksig \wi \gi}{\sumiksig \wi} 
= \frac{\sumiksig \fpkpy{\wi}}{\sumiksig \wi} 
= \fsenky{\ssetk}
\, ,
\label{eq:phisk}
\end{equation}
it is easy to see that 
$\gbar_{\kbin}\!=\!\purk\phisk$:
the net effect 
of background
is to dilute the overall bin-by-bin sensitivities
by a factor $\purk\!\le\!1$,
with respect to that computed
from signal events alone.~The 
same is also true for the 
bin-by-bin contributions to information,
which can be written as:
\begin{equation}
\setlength{\abovedisplayskip}{\abovedisplayskip-2mm}
\setlength{\belowdisplayskip}{\belowdisplayskip-2mm}
\Ify
= \frac{1}{(\Dy)^2}
= \sumkk \nsetk \gbar_\kbin^2 
= \sumkk \nsetk (\purk\phisk)^2 
= \sumkk \purk (\ssetk \phisk^2)
\, .
\label{eq:infoyksum3}
\end{equation}

For simplicity,
I will assume $\weighti\ofyder\!=\!\!1$
for all signal and background events
in the following.
This implies that
$\gbar_\kbin\!=\!(\sumik\!\gi)/\nsetk$
and 
$\phisk \!=\!(\sumiksig\!\gi)/\ssetk$
in the rest of this paper.

\newcommand{\ievto}{{\ievt_1}}
\newcommand{\ievtt}{{\ievt_2}}

\paragraph{An ideal measurement with an ideal detector, 
and
a realistic analysis
with a limited detector}
In my previous paper~\cite{bib:chep2018av},
I had shown that the optimal partitioning 
in a 
fit of~$\ytru$
consists in 
separating events into bins 
with different values of
the \mbox{bin-by-bin sensitivity~$\senky{\nset{k}}$}.
Event-by-event sensitivities
make it possible to 
go to 
a much finer granularity.

If only two selected events $\ievto$ and $\ievtt$ 
are expected,
the ``information inflow''~\cite{bib:vdbos}
in keeping them 
in separate one-event bins,
rather than mixing them together 
in a single two-event bin,
\begin{equation}
\setlength{\abovedisplayskip}{\abovedisplayskip-2mm}
\setlength{\belowdisplayskip}{\belowdisplayskip-2mm}
\Delta\Ify 
= \gvar_\ievto^2 \!+\! \gvar_\ievtt^2 \!-\! 2\left(\!\frac{\gvar_\ievto\!+\!\gvar_\ievtt}{2}\!\right)^{2}
\!= \frac{1}{2}(\gvar_\ievto \!-\! \gvar_\ievtt)^2 \, ,
\end{equation}
is zero if $\gvar_\ievto$ and $\gvar_\ievtt$ are equal,
whereas it is strictly positive if they are different.
In other words, 
in the ``ideal'' case where
all true values of the event-by-event 
sensitivities~$\gi$ were known,
the optimal way 
to measure $\ytru$
would be a fit 
of the one-dimensional distribution of~$\gvar$.
The maximum information $\Ify\ideal$ 
that is theoretically achievable 
in this ideal case is~simply
\begin{equation}
\setlength{\abovedisplayskip}{\abovedisplayskip-2mm}
\setlength{\belowdisplayskip}{\belowdisplayskip-2mm}
\Ify\ideal 
\!=\! \frac{1}{(\Dy\ideal)^2}
\!=\! \sum_{\ievt=1}^{\ntot} \gi^2 
\!=\! \sum_{\ievt=1}^{\stot} \gi^2 
\, ,
\label{eq:ifyideal}
\end{equation}
where the sum over
all $\ntot\!=\!\stot\!+\!\btot$ events 
includes $\stot$ signal and $\btot$~background events,
but the contribution from the latter is~0
because they have $\gi\!=\!0$ 
as described in Eq.~\ref{eq:gisb}.

\newcommand{\fipeff}{\fip_{\!\text{efS}}}
\newcommand{\fipsha}{\fip_{\!\text{shS}}}
\newcommand{\fippur}{\fip_{\!\text{shB}}}
\newcommand{\fipanl}{\fiph} 
\newcommand{\fipdet}{\fiph^{\text{(max)}}}
\newcommand{\Ifyana}{\Ify}
\newcommand{\Ifydet}{\Ify^{\text{(max)}}}
\newcommand{\fipacc}{\fip_{\!\text{ACC}}}
\newcommand{\fipall}{\fip_{\!\text{ALL}}}

As in Ref.~\cite{bib:chep2018av},
I suggest to 
evaluate the quality
of a measurement 
using the ``Fisher Information Part'', a
dimensionless scalar metric in [0,1],
defined as the ratio
between the information
which was actually achieved,
in Eq.~\ref{eq:infoyksum3},
and that achievable in an ideal case,
in~Eq.~\ref{eq:ifyideal}:
\newcommand{\ds}{}
\begin{equation}
\setlength{\abovedisplayskip}{\abovedisplayskip-2mm}
\setlength{\belowdisplayskip}{\belowdisplayskip-2mm}
\displaystyle
\fipanl
\!=\! \displaystyle
\frac{\,\Ify\hphantom{\ideal}\hspace*{-2mm}\,}{\,\Ify\ideal\,} 
\!=\! \displaystyle\frac
{\ds\sumkk \nsetk \gbar_{\kbin}^2}
{\ds\sum_{\ievt=1}^{\stot}\gi^2}
\!=\! \displaystyle\frac
{\ds\sumkk \ssetk \purk\phisk^2}
{\ds\sum_{\ievt=1}^{\stot}\gi^2}
\, .
\label{eq:fipgi}
\end{equation}
In Eq.~\ref{eq:fipgi},
the numerator is a sum over bins, 
based on metrics derived 
from the $\nsel\!=\!\sumkk\nsetk$ 
selected events in those bins
(where $\nsel\!=\!\ssel\!+\!\bsel$,
including $\ssel$ signal and 
$\bsel$ background events),
while the denominator 
is a sum over~the $\stot$ 
signal events in a given data sample.
The main difference between this metric 
and that I had previously 
presented~\cite{bib:chep2018av} is that
in the past 
I only used FIP to evaluate
the quality of event selection and 
signal-background discrimination
in a fit with a given binning,
while now I redefine it 
to also evaluate 
the quality of the binning.

FIP is a valuable metric
because it is simple to use and interpret
both qualitatively and quantitatively,
in statistically-limited measurements:
qualitatively, in that 
an analysis should be optimized 
to achieve the highest value of $\fip$;
quantitatively, in that 
its numerical value is proportional 
to $1/\Dy^2$,
where $\Dy$ is the statistical error
on the measurement.
Another useful feature 
is that,
since it is a ratio between 0 and 1,
FIP can be decomposed~as~the product 
of several independent metrics 
which are also ratios between 0 and 1.
In particular,
I propose to distinguish between three effects
which can result in information loss,
and I decompose
$\fipanl$
in Eq.~\ref{eq:fipgi}
as the product of three 
ratios, each taking values 
between 0 and~1:
\renewcommand{\ds}{}
\begin{equation}
\setlength{\abovedisplayskip}{\abovedisplayskip-2mm}
\setlength{\belowdisplayskip}{\belowdisplayskip-1.2mm}
\displaystyle
\fipanl
=
\displaystyle\frac
{\ds\sumkk \ssetk \purk\phisk^2}
{\ds\sum_{\ievt=1}^{\stot} \gi^2}
=
\displaystyle\frac
{\ds\sum_{\ievt=1}^{\ssel} \gi^2}
{\ds\sum_{\ievt=1}^{\stot} \gi^2}
\!\times\! 
\displaystyle\frac
{\ds\sumkk \ssetk \phisk^2}
{\ds\sum_{\ievt=1}^{\ssel} \gi^2}
\!\times\! 
\displaystyle\frac
{\ds\sumkk \ssetk \purk\phisk^2}
{\ds\sumkk \ssetk \phisk^2}
=
\fipeff\!\times\!\fipsha\!\times\!\fippur
\, .
\label{eq:fipgi3}
\end{equation}
The symbols
$\fipeff$, $\fipsha$ and $\fippur$
denote 
that these ratios 
represent effective 
measures~of
signal efficiency and of
signal and background ``sharpness''.
The concept of sharpness
(also known as "resolution",
a more familiar term in HEP)
describes the effectiveness at separating 
different categories of events from one another.
$\fipeff$ is an information-weighted signal 
selection efficiency,
describing
the loss of information
in rejecting some events:
it is the ratio between 
the $\ssel$ selected and $\stot$ total signal events,
where each event is weighted
by its 
information contribution $\gi^2$,
the square of its event-by-event sensitivity.
$\fipsha$
measures the sharpness at resolving
selected
signal events with different sensitivities $\gi$,
i.e. at partitioning them 
into different bins of the distribution fit,
$\ssel\!=\!\sumks\ssetk$:
it is the ratio of the information
achieved in the chosen binning $\Kbin$,
to that theoretically achievable
if it were possible to partition 
signal events
according to the true value $\gi$ 
of their sensitivity to $\ytru$.
$\fippur$ is an
information-weighted signal selection purity,
describing the loss of information
due to an imperfect background rejection,
in a given binning scheme $\Kbin$:
it too measures a ``sharpness'',
that at resolving 
background events (with $\gi\!=\!0$)
from 
signal events (of any sensitivity $\gi$).

\newcommand{\artwo}[2]{\begin{array}{c}\mathlarger{#1}\\[0.1cm]\boxed{\mathsmaller{#2}}\end{array}}
\begin{figure*}[tb]
\vspace*{-6mm}
\begin{center}
\begin{tikzcd}[row sep=huge, column sep=huge]
  \artwo{\sALL, \gi, \ddi}{\Ify\idealALL \!=\! \sum_{\ievt=1}^{\sALL}\gi^2} 
  \arrow[d, "\fipacc"] 
  &
  \hspace*{-0.5cm}
  \begin{array}{c}
    \\[0.3cm]
    {\fipall = \fipacc \times \fipanl} \\[0.1cm]
    {\fipanl = \fipeff \times \fipsha \times \fippur} \\[0.1cm]
    {0 \le \fipanl \le \fipdet \le 1}\\[-1.0cm]
  \end{array}
  \\[-0.1cm]
  \artwo{\stot, \gi, \ddi}{\Ify\ideal \!=\! \sum_{\ievt=1}^{\stot}\gi^2}
  \arrow[d] 
  \arrow[r, "\fipeff"] 
  \arrow[ddr, "\fipanl = \frac{\raisebox{1pt}{$\Ifyana$}}{\raisebox{-5pt}{$\Ify\ideal$}}" near start] 
  \arrow[dd, bend right=80, "\fipdet = \frac{\raisebox{1pt}{$\Ifydet$}}{\raisebox{-5pt}{$\Ify\ideal$}}" near start] 
  &
  \artwo{\ssel, \gi, \ddi}{(\Ify \!=\! \sum_{\ievt=1}^{\ssel}\gi^2)}
  \arrow[d, "\fipsha"] 
  \\[-0.1cm]
  \artwo{\stot, \phisx, \ddi}{(\Ify \!=\! \int \ssetx \phisx^2 d\xvec)}
  \arrow[d]
  &
  \artwo{\ssel, \phisk, \ddi}{(\Ify \!=\! \sumkk \ssetk \phisk^2)}
  \arrow[d, "\fippur"] 
  \\[-0.1cm]
  \artwo{\stot, \phisx, \purx}{\Ifydet \!=\! \int \ssetx \phisx^2 \purx d\xvec}
  &
  \artwo{\ssel, \phisk, \purk}{\Ifyana \!=\! \sumkk \ssetk \phisk^2 \purk}
\end{tikzcd}
\vspace*{-4mm}
\end{center}
\caption{
Graphical representation of FIP metrics and of their inter-relationships.
For each of the seven scenarios considered,
the number of signal events used 
and the resolution on signal sensitivity $\gi$
and on signal/background classification $\ddi$ 
(where $\ddi\!=\!1$ and $\ddi\!=\!0$
for true signal and background events, respectively)
are reported, 
as well as the 
information $\Ify$ 
which can be achieved from the measurement.
The main FIP metric discussed in this paper 
is $\fipanl\!=\!\Ifyana/\Ify\ideal$,
which is 
the product of $\fipeff$, $\fipeff$ and~$\fippur$. 
While theoretically $\fipanl$ is a metric 
in [0,1],
for a realistic detector $\fipanl\le\fipdet$. 
\vspace*{-5mm}
} \label{fig:fipchain}
\end{figure*}
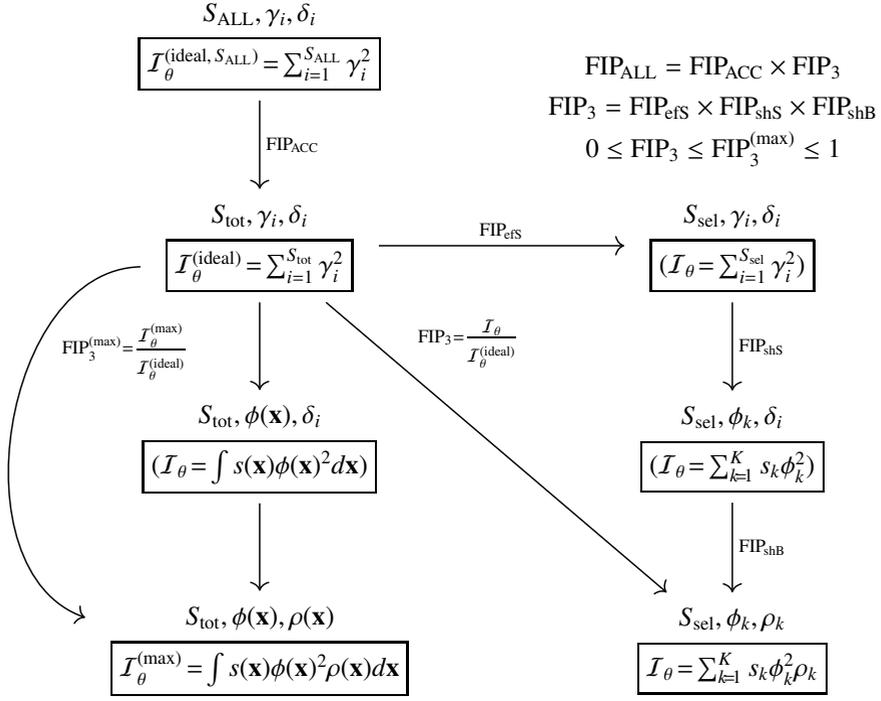

While I suggest
the use of $\fipeff$, $\fipsha$ and $\fippur$ 
as figures of merit
for the information-weighted efficiency
and signal and background sharpness
achieved by
the final analysis stage of a measurement,
it is 
important to point out that,
for all these three effects,
the maximum achievable figure of merit
with a realistic detector
may be lower than 1 
even if~the best possible analysis method is used.
Some loss of information may in fact 
be inevitable
given the limitations of the detector,
but also those of the computing and data processing chain
which precedes the final analysis stage of a measurement.
This is shown schematically 
in Fig.~\ref{fig:fipchain}.
To start with,
the $\stot$ signal events
in a final analysis sample
may be fewer than
the $\sALL$ signal events
produced in beam collisions
in the given data taking period,
because of detector acceptance, 
trigger decisions and 
preselection cuts:
this may be taken into account
by another ratio $\fipacc$,
analogous to $\fipeff$ and lower than 1,
by which the analysis-level $\fipanl$
should be multiplied to obtain 
the overall $\fipall$ metric for the measurement.
$\fippur$, or $\fipsha$, respectively,
may be lower than 1
because the limited resolution of the detector
mixes together signal events with different sensitivities $\gi$,
or mixes together
signal events and background events,
respectively,
making them experimentally indistinguishable.
For a real detector,
even the best possible analysis method
can at most try to determine,
at each point $\xvec$ of the observable phase space,
the average local sensitivity of signal events 
$\phisx\!=\!\gbar_{\text{Sig}}(\xvec)\!=\!\senky{\ssetx}$
and the average local purity
$\purx \!=\! (\ssetx)/(\ssetx\!+\!\bsetx)$
that the detector resolution
effectively establishes.
In these expressions,
$\ssetx$ and $\bsetx$ indicate 
the differential distributions
of signal and background events in $\xvec$-space,
with $\int\!\ssetx d\xvec \!=\! \stot$ 
and $\int\!\bsetx d\xvec \!=\! \btot$.

While the framework I propose
describes the general case
where signal events 
have different sensitivities $\gi$ to $\ytru$
and are thus not all equivalent to one another,
it also describes~a~much simpler case
where signal events all have 
the same sensitivity $\gi$,
namely the measurement 
of a total signal cross section $\sigs$.
In this case,
which I 
discussed in Ref.~\cite{bib:chep2018av},
the only challenge 
is the classic binary classification problem
of signal-background discrimination
in the presence of strictly dichotomous true categories.
As there is no need 
to resolve signal events from one another,
$\fipsha$ is always 1 in this case.
If $\sigs$ is measured 
by a counting experiment (i.e. using~a single bin),
$\fipanl$ reduces to
to $\fipo\!=\!\effs\gpur$~\cite{bib:chep2018av},
the product of the 
global signal selection efficiency $\fipeff\!=\!\effs$
and purity $\fippur\!=\!\gpur$,
a metric that has been widely used in HEP
already since 
the late 1990s~\cite{bib:yellow1996,bib:valassi-phd,bib:opal-ww-1998,bib:aleph-lemaitre-2000,bib:delphi-ww-2000}.
Another common way to measure $\sigs$ is 
the fit of a scoring classifier distribution:
examples include fits of Neural Network 
or Rarity distributions
at LEP~\cite{bib:aleph161}
and fits of Boosted Decision Trees
at the Tevatron~\cite{bib:d0-singletop-2008,bib:cdf-singletop-2009} 
and LHC~\cite{bib:cms-singletop-2011}.
In this case,
$\fipeff\!=\!1$ because all pre-selected events 
are included in the fit,
while $\fippur$ 
reduces to 
$\fipt\!=\!(\sumks 
\ssetk\purk)/(\sumks \ssetk)$~\cite{bib:chep2018av},
because $\gi$ is the same for all signal events.

\paragraph{Monte Carlo weight derivative regression}

To optimize the measurement of~$\ytru$
from~a sample of $\ntot$ events,
it would 
then
be enough to know a single
property of all events,
\mbox{their sensitivity~$\gi$ to~$\ytru$}.
The fit of the one-dimensional distribution of $\gi$
would provide optimal partitioning 
and~background rejection, and
achieve the minimum statistical 
error $\Dy\ideal$.
The challenge to address is that,
while $\gi$ can  be  computed  for  MC events,
$\gi$ is not available 
for real data.
The practical strategy I suggest 
is to train
a regressor $\gregi$
of $\gi$ on MC events, 
i.e. a regressor of the MC 
\mbox{weight derivatives $\senky{\wi}$}
computed from the generator-level 
properties $\xvecgeni$
of MC events,
and use it
to fit $\ytru$ from the 
one-dimensional distribution 
of $\gregxi$ on data events,
computed from their detector-level
properties $\xveci$.
I refer to this approach
as ``Weight Derivative Regression''~(WDR).

In such a crude form,
this method is probably
of little applicability
in many practical~situations,
and more refined variations
should be used to overcome
some of its limitations.
The main issue
is 
that the MC weight derivatives
$\senky{\wi}$ depend 
on the value~$\ytruder$ of $\ytru$
at which they are computed:
this dependency may be 
weak in fits of particle couplings,
but is certainly strong 
in fits of particle masses.
It may be necessary to compute these
derivatives at more than one value 
of $\ytruder$,
and possibly train more than one regressor,
using them to measure~$\ytru$ from a multi-dimensional fit.
A separate binary classifier for background
rejection may also be useful,
especially to handle systematic errors.
A more detailed discussion
of the limitations of this method,
and practical examples of its use,
\mbox{are foreseen for later publications}.

I stress that
the method I suggest 
has clear similarities with,
and was strongly inspired by,
the ``Optimal Observables'' (OO)
approach~\cite{bib:optobs92,bib:optobs93-tau,bib:optobs93-tgc,bib:optobs05}.
There is, however, an important difference,
which schematically is the following:
the WDR method
consists in 
computing the true sensitivity $\gi$ 
of each MC event~$\ievt$\
from its generator-level
properties $\xvecgeni$,
and training the regressor 
$\gregi\!=\!\gregxi$ 
against these true $\gi$,
to obtain an estimate 
$\gregx$
of the functional dependency 
of the local average sensitivity
$\gbarx\!=\!\phisx\purx$
on the detector-level
properties $\xvec$ for real data events;
the OO method 
approximately consists, instead,
in analytically computing 
the functional dependency of $\gi$ on $\xvecgeni$,
and applying that same functional dependency 
on the observed $\xvec$
to obtain an estimate of~$\gbarx$ 
for real data events.
As a consequence, the results that can be obtained
through the OO method
are significantly degraded 
by the effect of the experimental 
detector resolution, which is
not properly accounted for.

\newcommand{\sumit}{\sum_{\ievt\!=\!1}^{\ntot}}

The regressor $\gregi\!=\!\gregxi$ 
of the sensitivity $\gi$
may be implemented 
in many different ways.
Selecting a specific algorithm
essentially means choosing two things:
the parametrization of the $\gregx$ function,
and the metric 
to use for
training the regressor.
As 
in Ref.~\cite{bib:chep2018av},
I focus on
Decision Tree (DT) algorithms~\cite{bib:cart},
and I suggest that
the maximization of $\fipanl$ 
should be used both
for evaluating the measurement
and for training the regressor.
In a DT,
the space~of detector-level event properties $\xvec$
is split into $\nbin$ disjoint nodes, 
such that $\gregx\!=\!\gregk$ 
is a constant in each node $\kbin$.
Taking into account
that each node of the tree 
may be used as a bin in the fit,
the goal is
to split all $\nsel\!\!=\!\!\ntot$ events 
in the training sample 
into $\nbin$ nodes/bins, 
with $\nsetk$ events 
in node/bin $\kbin$,
so as to maximize $\fipanl$ in Eq.~\ref{eq:fipgi}.
It is 
extremely interesting to see that
this is equivalent 
to using a much more common criterion,
the minimization of
the Mean Squared Error (MSE).
It is easy to prove, in fact,
that the MSE can be decomposed as follows,
\newcommand{\MSEsha}{\MSE_{\text{sha}}}
\newcommand{\MSEval}{\MSE_{\text{cal}}}
\vspace*{-2mm}
\begin{align}
\MSE 
\!=\!
\frac{1}{\ntot}\!\sumit(\gregi\!-\!\gi)^2
& \!=\!
\frac{1}{\ntot} \!
\left[ \sumkk \!\nsetk \!\left(\gregk \!-\! \gbark \right)^2 \right]
\!+\!
\frac{1}{\ntot} \!
\left[ \left(\! \sum_{\ievt=1}^{\ntot} \gi^2 \!\right) \!-\! \left(\! \sumkk \nsetk \gbark^2 \!\right) \right]
\label{eq:mse}
\\ & 
\!=\!
\frac{1}{\ntot} \!
\left[ \sumkk \!\nsetk \!
\left(\gregk \!-\! \gbark \right)^2 \right]
\!+\!
\frac{1}{\ntot} \!
\left[ \Ify\ideal \!-\! \Ify \right]
\!=\!
\MSEval\!+\!\MSEsha
\nonumber
\, , 
\\[-7mm]\nonumber
\end{align}
where the ``calibration'' 
$\MSEval$ is~0 by construction 
in training the DT,
as $\gregk$ is defined as the average
sensitivity $\gbark$ of the MC events in node $\kbin$,
while the ``sharpness'' 
$\MSEsha$ is minimized when $\fipanl$ 
(or more precisely $\fipsha\times\fippur$,
as $\nsel\!\!=\!\!\ntot$)
is maximised,~because
\begin{equation}
\setlength{\abovedisplayskip}{\abovedisplayskip-2mm}
\setlength{\belowdisplayskip}{\belowdisplayskip-2mm}
\left( 1 \!-\! 
\frac{\ntot\times\MSE_{\text{sha}}}{\Ify\ideal} \right)
\!=\!
\frac{\Ify}{\Ify\ideal} 
\!=\!
\fipanl
\!=\!
\fipsha\times\fippur
\, .
\end{equation}
For other algorithms,
such as Neural Networks,
where implementing FIP maximization
is not as easy as in a DT,
minimizing MSE is probably 
still a sensible training criterion.
\vspace*{-1.5mm}

\section{Learning from others: probabilistic metrics in Meteorology}
\label{sec:brier}
\vspace*{-1.5mm}

I now take 
a step backwards 
to consider the more general perspective 
of evaluation and training metrics
in different scientific domains.
The reason why metrics like FIP and MSE
are relevant to HEP parameter fits
is that they capture 
their most characteristic feature,
the simultaneous use
of disjoint event partitions
to derive a measurement of $\ytru$
which is effectively 
a combination of the measurements
performed in these individual partitions.
It should~be~noted~in passing that 
most of the ideas in this paper 
are relevant for both binned and unbinned
fits,
even if their applicability 
is more obvious
in the case of 
binned fits.
In my previous study~\cite{bib:chep2018av},
I noted
that event partitioning
is largely unaccounted for 
by the evaluation metrics commonly
used in 
Medical Diagnostics (MD),
Information Retrieval~(IR)
and Machine Learning (ML).
Further research led me to understand
two things: first,
that a key point is the 
categorization~\cite{bib:caruana2004,bib:ferriorallo2004,bib:wu-flach-2007,bib:ferriorallo2009}
of performance metrics
into three~\mbox{distinct} families,
namely threshold, ranking
and probabilistic metrics;
and, second, that
MD, IR and ML mainly
focus on binary classification 
problems described by
threshold and ranking metrics, whereas 
HEP parameter fits
require probabilistic metrics,
which are 
widely used
for regression problems
in domains such as Meteorology and Climatology, 
or Medical Prognostics. 

Threshold metrics are relevant in 
classification problems
where all events are 
assigned 
to a 
signal or background category
by a discrete binary classifier. 
This includes the case
when the operating point 
of a scoring classifier
is chosen on its 
ROC~\cite{bib:peterson-roc-1953,bib:tanner-swets-1954,bib:peterson-roc-1954-pgit4,bib:tanner-swets-1954-pgit4,bib:vanmeter-middleton-1954-pgit4,bib:swets-tanner-birdsall-1955,bib:egan-1956,bib:swets-tanner-birdsall-1961,bib:birdsall-thesis-1973}
curve (for instance based on a cost matrix),
a popular approach in 
MD~\cite{bib:lusted1960,bib:lusted1968,bib:lusted1971,bib:metz-goodenough-1973,bib:metz-starr-1975,bib:mcneil-1975,bib:metz-roc-1978,bib:lusted1984}.
Classifiers are evaluated from the 
four~event counts 
in a two-by-two 
confusion matrix, 
namely True/False Positives/Negatives.
The simplest threshold metric 
is accuracy,
which is widely used, 
but is known to have severe limitations,
in both MD~\cite{bib:swets-1979-imaging,bib:swets-1988}
and ML~\cite{bib:ml-spackman-1989,bib:bradley,bib:provost-1997,bib:provost-1998,bib:fawcett-roc-2006}.
A popular threshold metric in 
IR~\cite{bib:ir-kent-1955,bib:ir-cleverdon-1962,bib:ir-swets-1963,bib:ir-cleverdon-1965,bib:ir-rijs1974,bib:ir-rijs1979,bib:ir-manning}
is the F1 score:
this is based on precision and recall,
which in HEP are known as
purity $\gpur$ and efficiency $\effs$.
In HEP, 
threshold metrics
are especially useful in counting experiments:
examples include cross section measurements by counting,
where the relevant metric 
is $\fipo\!=\!\effs\gpur$, as discussed,
but also searches for new 
physics~\cite{bib:punzi,bib:cousins,
bib:cowan2011,bib:higgsml}
that are not based on distribution fits.
An interesting way 
to compare
different threshold metrics
is to study
their symmetries and 
invariances~\cite{bib:sokolova,bib:luque}.
A fundamental feature of 
HEP measurements,
in particular, 
is the irrelevance 
of the True Negatives count,
i.e. of the number of rejected background events:
in this respect, HEP is more similar to IR
than it is to MD,
as I briefly discussed in Ref.~\cite{bib:chep2018av}.

Ranking metrics are 
relevant in 
classification problems
where all events 
are assigned a score $\dvar$
by a scoring classifier,
representing their probability 
to belong to the signal category.
Events can then be ranked 
by their score,
which is especially important
if some prioritization is needed.
Ranking metrics such as
precision 
for a fixed number of retrieved documents, 
or a fixed fraction of all available documents,
are often used in
IR~\cite{bib:ir-trec3-taguesutcliffe-1995,bib:ir-trec3-appendixa-1995,bib:ir-trec2-harman-1995,bib:ir-hull-1993}.
The most commonly used 
ranking metric
is however the Area Under the ROC Curve 
(AUC),
which~is~popular in 
MD~\cite{bib:green1964,bib:green-swets-1966,bib:goodenough-1972,bib:bamber1975,bib:hanley-auc-1983}
because it represents
``the probability that a randomly chosen diseased subject
is correctly ranked with greater suspicion
than a randomly chosen non-diseased subject''. 
The AUC is however known to have severe limitations
for both MD~\cite{bib:greiner2000,bib:zhou2002,bib:ray2010,bib:hajian2013} 
and ML~\cite{bib:adams-hand-1999,bib:drummond-2000a,bib:drummond-2006,bib:davis-goadrich-2006,bib:saito,bib:he-garcia-2009}.
Ranking metrics 
are an active area of research in 
ML~\cite{bib:clemencon-vayatis-2007,bib:clemencon-lugosi-2008,bib:rudin2018},
which was also investigated in HEP~\cite{bib:higgsml}.
In my opinion~\cite{bib:chep2018av}, 
however,
ranking metrics, and in particular the AUC, 
are~largely irrelevant in HEP measurements:
while threshold metrics are needed in 
counting experiments,
for distribution fits
one should use metrics
describing event partitioning,
not event ranking.
In a cross section fit 
from the distribution
of a scoring classifier $\dvar$,
for instance,
a metric like $\fipt$ is relevant
because it describes 
the fit as a combination
of measurements from subsets of events
with different values of~$\dvar$,
independently of which 
event subset
has a higher score.

A related challenge in HEP distribution fits
is that signal events are not all equivalent~to 
one another, 
as they have different sensitivities $\gi$.
Research on metrics 
for~non-dichotomous evaluation
has been active 
on non-binary gold standards in
MD~\cite{bib:pencina-2004,bib:obuchowski-2006,bib:lambert-2008}, 
on graded relevance assessment in
IR~\cite{bib:ir-jarvelin-2000,bib:ir-jarvelin-2002a,bib:ir-jarvelin-2002b} and 
on cost-sensitive classification in
ML~\cite{bib:turney-1994,bib:drummond-2000b,bib:zadrozny-2001,bib:elkan-2001,bib:zadrozny-2003,bib:fawcett-rociv-2006},
involving threshold, ranking 
and probabilistic metrics,
and even discussing the issue
of the calibration of probabilistic 
classifiers~\cite{bib:calib-zadrozny-2001c,bib:guo2017}.
In my opinion, however, a more appropriate
solution for HEP distribution fits
may come
from probabilistic metrics in other domains.

Probabilistic metrics are relevant 
in classification and regression problems
where the comparison of a
predicted property of an event
to its true value 
has a probabilistic interpretation.
Verification scores of forecasts in
Meteorology and Climatology~\cite{bib:brier1950,bib:sanders1963,bib:murphy1973,bib:licht1977,bib:mason1982,bib:murphy-winkler-1987,bib:wmo-svslrf-1992},
such as MSE and the closely related Brier score,
are typical 
probabilistic metrics.
Similar metrics are also used
for the evaluation of patient health predictions
in Medical 
Prognostics~\cite{bib:spiegelhalter,bib:harrell-lee-mark}.
In both cases,
the quality of forecasts is assessed
by comparing a forecast probability
of a future weather event, or of a future disease,
to the relative frequency 
which is eventually observed for that event.
Partitioning 
is an essential component of this approach:
for instance,
ten different forecast groups may be studied, 
each covering a 10\% probability range,
with the third group including 
days (or patients), 
with a 20 to 30\% 
probability of rain
(or of survival after 5 years, respectively).
A good forecast is one with two features:
first, reliability or calibration, 
i.e. the actual fraction of rainy days
must be $\sim$25\% for forecasts
in the~20--30\% range;
second, sharpness or resolution, 
i.e. it must be able to 
distinguish between
days with a $\sim$25\% probability 
and days with a $\sim$75\% probability of rain.
As discussed 
in Sec.~\ref{sec:fits},
probabilistic metrics 
like MSE,
and the concepts of sharpness and calibration
of a regressor are also 
relevant 
to describe
HEP parameter fits:
the decomposition in Eq.~\ref{eq:mse}
was, in fact, copied from
that of the Brier score into 
a calibration and a sharpness term 
in Meteorology~\cite{bib:sanders1963}.

\vspace*{-1.5mm}
\section{Outlook and conclusions}
\label{sec:concl}
\vspace*{-1.2mm}

I have described a mathematical framework
to evaluate
HEP parameter fits,
and suggested~a
MC Weight Derivative Regression 
approach
to optimize them.
Data analysis methods are similar 
across scientific domains,
and HEP can learn a lot from others;
but different problems require different metrics,
and it is important to select from other domains
the tools that make sense 
for us.
I pointed out
in particular
that ranking metrics like the AUC,
a standard practice
in Medical Diagnostics,
are of limited relevance for HEP,
while probabilistic metrics like the MSE
and the concepts of 
calibration and sharpness,
commonly used in Meteorology,
are directly applicable in our field.
I have not discussed systematic errors,
or searches for new physics
based on distribution fits,
but I hope that this work can stimulate
research in that direction.
Further details on this work
are available in the slides 
of the CHEP2019 talk~\cite{bib:avtalk}
described in this paper.
A more detailed article
is also planned for the future.

\end{document}